# The current unbalance in stacked REBCO tapes — simulations based on a circuit grid model

Rui Kang[1,*], Juan Wang[1], Ze Feng[1], Qingjin Xu[1]

[1] The Accelerator Division, Institute of High Energy Physics, 100049 China
[*] kangrui@ihep.ac.cn

**Abstract**

Unlike low-temperature superconducting cables, there is so far no perfect solution for REBCO coated conductors to form a fully transposed high-current cable. Every REBCO cable concept must import a stack of tapes to achieve an operating current as high as tens of kiloamperes. The stacked REBCO tapes, no matter whether they are twisted or not, however, have a nature of non-transposing and therefore could result in current unbalance. In this manuscript, the current unbalance and the related electrical characteristics of a cable made of 40 stacked REBCO tapes are studied with an electrical circuit simulation. The differences in splice resistances and tape inductances that are both related to the non-transposed structure of a REBCO stack are considered. Results show that for a 40 cm long termination, a proper method to keep the contact resistivity between each tape and the copper termination around 10 nΩ·m is crucial to totally avoid current unbalance lowering the cable's performance. Surprisingly, the inter-tape current transfer is found to be able to further exacerbate local high current though it does make the overall distribution more balanced. The inductance difference induced current unbalance is only important if local defects exist at long REBCO tapes, which on the other hand can be cured by good inter-tape current transfer. For a fast-charging rate of 1 kA/s, the inter-tape contact resistivity should also be low to a level of 10 nΩ·m to ensure a short current transfer length of around 1 m.

## 1. Introduction

Transposition is usually mandatory for superconducting cables made of low-temperature superconductors (LTS) [1,2], which ensures low AC losses and balanced current distribution among the wires. However, for REBCO which is the most attractive high-temperature superconductor (HTS) at present, a fully transposed high-current cable seems not yet possible. In the last two decades, many REBCO cable concepts have been developed as reviewed in [3]. Among them, the only cable that fully transposes the HTS tapes is the Roebel cable. Despite the halved current carrying capability due to cut-out, a classic Roebel cable could only contain a very limited number of tapes. To obtain a high current with the Roebel structure, one zigzagged tape must be replaced with a stack of a few [4], which on the other hand leads to non-fully transposition. Indeed, most of the high-current REBCO cable concepts introduce a stack of tapes from the very beginning [5–11].

A stack of REBCO tapes has a nature of non-transposing. Twisted or not, the relative positions of the tapes in one stack would never change, as shown in figure 1. Such a feature is unfavorable for reducing the AC loss and achieving a balanced current distribution for REBCO cables [12]. As AC loss has been studied a lot, this manuscript focuses on the latter issue.

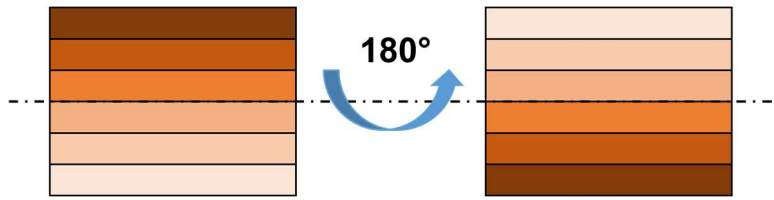

**Figure 1** A view of twisted stacked tapes. The relative position among tapes does not change.

As far as we know, there are generally three kinds of factors that may result in an unbalanced current distribution in a stack of REBCO tapes:

1) **The difference in inductances**.

The unequal inductances of the tapes in one stack are a direct consequence of being non-transposed, as calculated in [5,12]. The tapes at the center would have higher inductances than those at the top or bottom. When charging a coil wound with such a cable, the tapes at the center endure higher inductive voltages, so the current will preferentially path through the tapes at the bottom or top, until the current is close or even higher than the critical current and generates resistive voltages large enough to compensate the differences.

2) **The difference in the splice resistances**.

Even for transposed LTS cables, inhomogeneous contact resistances between the strands and the copper termination are still inevitable [13]. For a REBCO cable with stacked tapes, it would be worse. The non-transposition means the tapes have different opportunities to contact the splice. Tapes at the top or bottom could have direct contact with the splice, but those at the center can only indirectly transfer current through other tapes or solder at the edges. Besides, REBCO tapes are asymmetric and higher resistance is expected if current goes through the substrate side. This feature will further diversify the splice resistances of tapes. When a transport current is applied to a REBCO cable, the current will first distribute over those tapes with low splice resistances, until the resistive voltages are large enough to compensate for the differences. Although there are methods to achieve a low and uniform splice resistance for REBCO cables, inhomogeneity is still expected [14–16].

3) **The difference in critical currents**.

Though one important benefit of using a large current cable is to overcome the non-uniform critical current of a single conductor, it has been proved that the difference in critical currents could also cause unbalanced current distribution in tapes [17]. Note that the different magnetic fields each tape sees could also lead to different critical currents [11,15].

In practice, the importance of these factors may vary depending on the situation. For short sample measurements, the splice resistances are expected to dominate the current distribution in stacked REBCO tapes. In a large-scale coil wound with long cables of hundreds of meters, current unbalance in a stack of REBCO tapes is most likely generated by the inductance variation. Contribution from the other two factors could further worsen the situation.

Though the unbalanced current distribution could greatly limit the performance of a superconducting cable as reported in both LTS and HTS cables [2,5], compared to the great passion for superconducting cables, this topic is not often studied. As far as we know, for REBCO cables there are only several studies that discussed the unbalanced current caused by the different splice resistance [5,15,18–20]. In most of these studies, the tape numbers are too few or the current is too low that the results can hardly be imitated to high current cables. The only exception is the one Takayasu *et al* reported in 2016 [20], in which a stack with 40 tapes is studied. The study for the unbalanced current driven by the different critical currents has also been studied [11,15,17]. As for the difference in inductances induced current unbalance, no studies are reported yet for REBCO cables. It should be noted that the calculation of screening current in one REBCO film has the same nature as the inductance induced current distribution in a large cable. The difference is just the scale. For a coil with hundreds of turns and each turn containing tens of tapes, the electrical circuit description gives a much more simple solution. Another practical problem for us is that during the test of a Roebel structured REBCO cable (named X-Cable [11]), the cable performance is significantly lower than the sum of single tapes even considering the self-field and the test sample is frequently burnt at places near the termination. Considering these, we believe it is very interesting and necessary to have a further quantitative study on the current unbalance in stacked REBCO tapes and its influence on the cable's electrical characteristics.

In this manuscript, we try to study the current unbalance and the related electrical characteristics of a cable made of 40 stacked REBCO tapes with an electrical circuit simulation. Section II is a brief description of the electrical circuit grid model. A stack of 40 REBCO tapes is assumed as an object. Section III deals with the performance of short samples. As a start, the splice resistances in several different cases are reviewed. Then three kinds of terminations are compared, with different magnitudes of splice resistance as well as their differences. The current distribution, voltage, and joule heating are compared. Next, the current transfer among tapes at the cable part (outside the terminations) is considered. The existence of a copper shell and whether it is soldered into the termination are also studied. In section IV, the different inductances induced current unbalance is studied for a hypothetical coil wound with a 40-stacked tape cable. The effect of local defects and possible cure by inter-tape current transfer are also quantitatively discussed.

## 2. A circuit grid model for stacked REBCO tapes

A cable made of 40 stacked REBCO tapes is studied. The relevant parameters are summarized in Table 1. At each end of the cable, a copper block is connected to the tapes. The electrical characteristic of such a cable could be simulated with an electrical circuit model, as shown in figure 2. The topology refers to the second type of termination mentioned in section 3.1. Each REBCO tape is divided into three parts: two inside the copper block (referred to as inside parts) and one outside (referred to as outside part). The same length is assumed for the two inside parts, and each is discretized into *nt* elements. The outside part is discretized into *nc* elements. The resistance of each REBCO ($R_{sc}$) element is calculated by solving the following equations [21]:

$$\frac{1}{R_{sc}} = \frac{1}{\frac{E_c}{I_c}(\frac{I_{sc}}{I_c})^{n-1}} + \frac{1}{R_n} \quad (1)$$

$$E_c(\frac{I_{sc}}{I_c})^n = (I_s - I_{sc})R_n \quad (2)$$

In which $E_c$ is the electrical field criterion for critical current ($I_c$) of REBCO and is set at 1 µV/cm here. In this study, as the focus is on the difference of splice resistance and inductance that both are related to the non-transposition nature of stacked REBCO tapes, $I_c$ of REBCO is assumed a constant of 147 A, which is a typical value of a 4 mm wide REBCO tape at 77 K and self-field or at 4.2 K and high vertical field beyond 20 T. The power law index *n* is also assumed a constant of 30. $I_{sc}$ is the current in the REBCO film. $R_n$ is the homogeneous resistivity of the normal metal parts (copper, silver, and Hastelloy in this case) in a REBCO tape taken the value at 77 K. $I_s$ is the total current in one REBCO tape element. As the influence of temperature and magnetic field are both ignored, eventually the resistivity of each REBCO tape becomes a function of its current $I_s$, as shown in figure 3.

**Table 1.** Parameters of the REBCO tape

| Parameter | Value |
|---|---|
| Tape $I_c$ | 147 A |
| Tape width | 4 mm |
| Tape composition | REBCO: Hastelloy: Cu: Ag = 1:50:20:4 |
| Tape thickness | 0.075 mm |

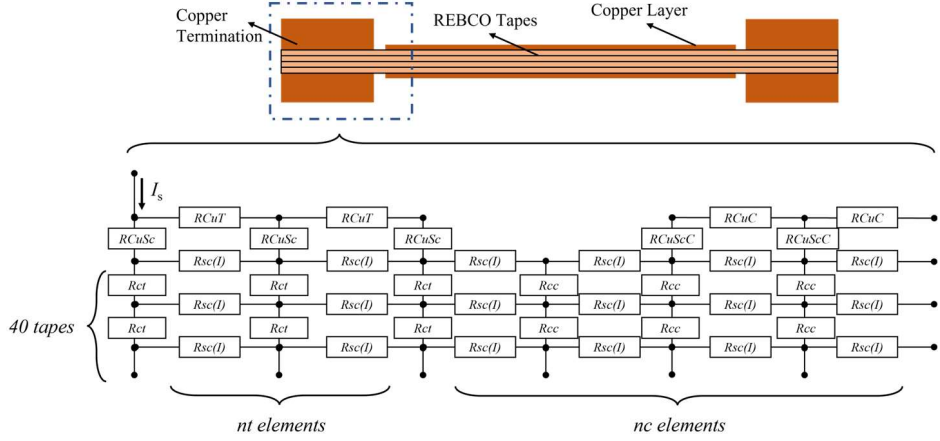

**Figure 2.** A schematic of the circuit grid model for stacked REBCO tapes.

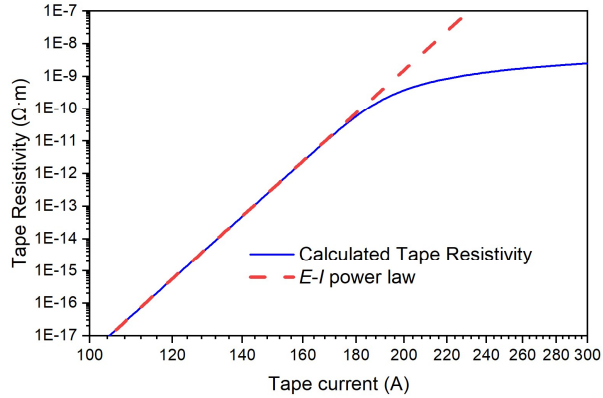

**Figure 3.** Tape Resistivity as a function of tape current. The blue solid line is calculated with equations (1-2) considering the conductivity of normal metals. The red dashed line is what the *E-I* power law gives.

Depending on the termination techniques, the REBCO tapes have different contact conditions with each other or with the copper termination, as later reviewed in section III. The former is defined by $R_{ct}$ and the latter is defined by $R_{CuSc}$. The two nodes of each REBCO element connect with the parallel REBCO or copper element with these two types of contact resistances, as also shown in figure 2. For the outside part, the contact resistance (if the inter-tape current sharing is considered) between REBCO tapes is defined by a variable $R_{cc}$. For an outside REBCO element, an inductor can also connect with the resistor in series. The mutual inductance between each two parallel REBCO elements is considered in such a case.

For the simulation of short testing samples, the calculation is stationary, and the inductors are neglected. The length of the outside part ($L_c$) is assumed 1 m and is discretized into 100 elements ($nc$=100). The length of each termination ($L_t$) is assumed 0.4 m and $nt$ is 40. As a result, each element has a length of 1 cm and it is proven a good balance between calculation efficiency and accuracy. For the hypothetical coil that is used to study the current unbalance caused by inductance difference, $L_t$ remains to be 0.4 m, but $L_c$ increases to 100 m. On the other hand, as the total number of inductances (self-inductance plus mutual inductances) that must be calculated is proportional to the square of tape numbers, it is found difficult to properly discretize the 40 tapes over 100 m. Consequently, in this part of the simulation, the same splice resistance is assumed for all tapes, and both *nc* and *nt* are set to 1. All the calculations are implemented with the electrical circuit module in the COMSOL software.

## 3. The current unbalance in short samples resulted from splice resistance.

3.1 The splice resistance

As far as we know, the termination techniques could be roughly divided into three types for REBCO cables. The first type is to only solder one tape directly to the copper termination and let the ceramic side face the copper. Then stack the remained tapes and solder them together, so the current must bypass the substrate and buffers when the solder layer is very thin at the side of the stack, as reported in [19]. Such a technique is apparently not suitable for a REBCO stack with many tapes but could be an extreme case to study. As is reported in [19], the contact resistivity between the copper block and the directly contacting REBCO tape is about 35 nΩ·m, but the value between two REBCO tapes ranges from 192 to 266 nΩ·m, almost an order of magnitude higher. The latter values are also consistent with a measurement of inter-tape resistivity reported in [22]. The second type is to insert the whole cable into the copper termination with a pre-machined channel, and then fill the void with solder, as used in [11,23]. This is also a straightforward way to get a rather modest splice resistance since the current could go through the side path. A quantitative estimation for the splice resistance in this kind of termination method could be achieved according to the inter-tape resistivity measured in stacked REBCO tapes soldered inside copper jackets [16]. The equivalent contact resistivity between each tape to the copper block (jacket) is found to be evenly distributed from 6.5 to 18 nΩ·m at 77 K and could be even lower to a quarter at 4.2 K. However, when a similar approach is applied to the test of our prototype of X-cable, the contact resistivity is at the level of 100 to 1000 nΩ·m, which could be due to the thicker solder layer as well as the higher resistivity of the solder with low melting temperature. There are also several dedicated methods that have been developed, for example staggered or folding-fan arrangement of the REBCO tapes to make sure each REBCO tape directly contact the copper block [14–16]. Surprisingly, these methods do not show significant improvements comparing with the second route. The measured values from the three references are respectively from 2.5 to 7.5 nΩ·m, 12 to 20.2 nΩ·m, and 10 to 100 nΩ·m.

Based on the above results, in this study three kinds of termination resistivity (Rt) are discussed:
1) High-Rt, only one REBCO tape directly contacts the copper block with a resistivity of 35 nΩ·m, the other tapes contact each other with the same resistivity of 229 nΩ·m.
2) Mid-Rt, every tape directly contacts the copper block, but the resistivity ranges from 10 to 1000 nΩ·m uniformly.
3) Low-Rt, every tape directly contacts the copper block, but the resistivity ranges from 12 to 20 nΩ·m uniformly.

The 40 tapes are numbered in ascending order according to the contact resistivity with the copper block. There are in total 13 cases studied, as summarized in Table 2.

**Table 2.** A summary of the simulated cases for short cables

| Case No. | Case Classification | Splice resistance | $R_{cc}$ | $R_{CuScC}$ |
|---|---|---|---|---|
| 1 | Without inter-tape current transfer outside the termination | High | \ | \ |
| 2 | | Mid | \ | \ |
| 3 | | Low | \ | \ |
| 4 | With inter-tape current transfer outside the termination | High | 1000 nΩ·m | \ |
| 5 | | High | 100 nΩ·m | \ |
| 6 | | Mid | 1000 nΩ·m | \ |
| 7 | | Mid | 100 nΩ·m | \ |
| 8 | | Mid | 10 nΩ·m | \ |
| 9 | | Low | 1000 nΩ·m | \ |
| 10 | With inter-tape current transfer outside the termination and having the copper layer | High | 1000 nΩ·m | 10 μΩ·m |
| 11 | | High | 1000 nΩ·m | 1 μΩ·m |
| 12 | | Mid | 1000 nΩ·m | 1 μΩ·m |
| 13 | With inter-tape current transfer outside the termination and having the copper layer. The copper layer is soldered inside the termination | High | 1000 nΩ·m | 10 μΩ·m |

3.2 The current unbalance resulted from splice resistance.

The first three cases focus on how the splice resistance alone influences the current distribution among a REBCO stack as well as the corresponding electrical characteristic, assuming the tapes are insulated with each other at the outside cable part. Figure 4 compares the current in the outside cable part in different tapes as a function of total current. Since current always prefers to pass through the tape with the minimum resistance, current unbalance is inevitable even for a case with very low but not totally uniform splice resistances. The difference is, in the low-Rt case, once the current in one tape is approaching its $I_c$, the tape resistance is soon comparable to the small splice resistance. Consequently, for the low-Rt case, when the total current is close to the sum of tape $I_c$, the current is almost evenly distributed over the 40 tapes and no tape hosts a current higher than its critical current. For the mid-Rt case, things already become a little bit different. When the total current is 6000 A, the current varies from 139 A to 152 A in different tapes. In the high-Rt case, at 6000 A there are still 5 tapes in which there is almost no current at all. On the contrary, there are 33 tapes in which the current exceeds the $I_c$ and the maximum is up to 190 A. It is also worth mentioning that in the low and mid-Rt cases, the highest current is not always in the tape with the minimum splice resistance after the first tape enters a saturated state.

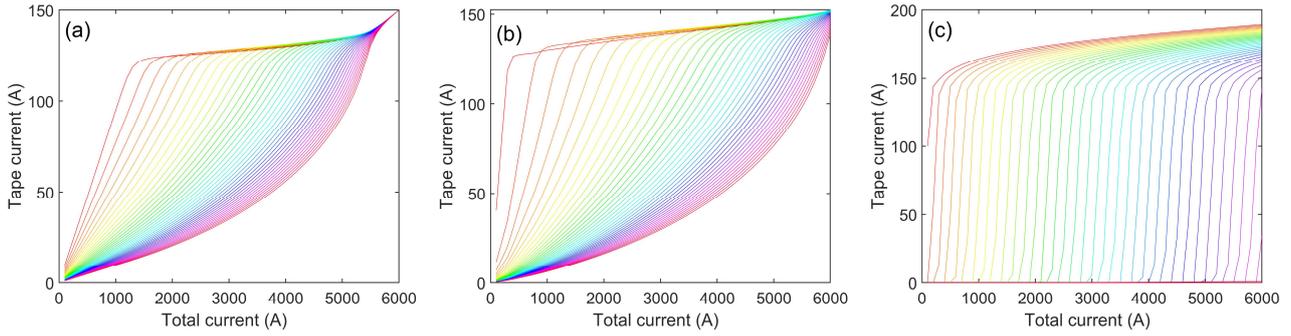

**Figure 4.** Current in each tape at the outside cable part as a function of the total current in the case of (a) low Rt, (b) mid Rt and (c) high-Rt. The 40 tapes are represented by colors from red (for tape No. 1) to purple (for tape No. 40).

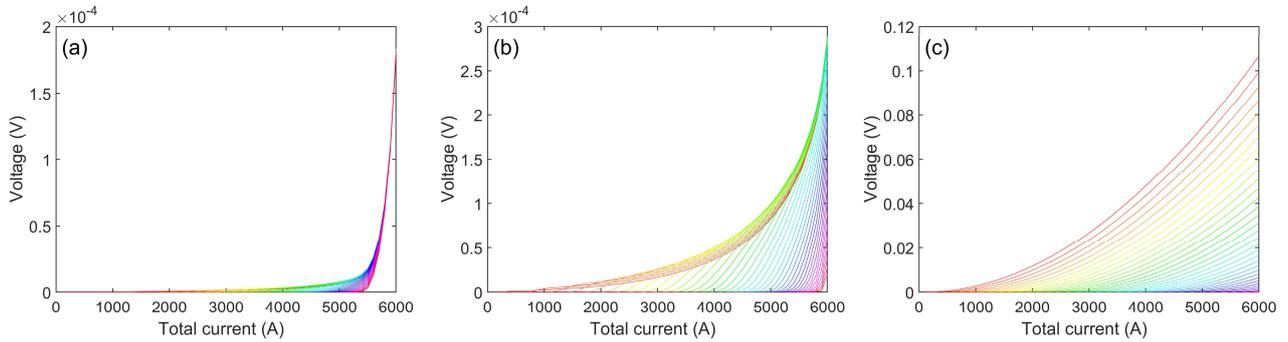

**Figure 5.** Voltage in each tape at the outside cable part as a function of the total current in the case of (a) low-Rt, (b) mid-Rt, and (c) high-Rt. The 40 tapes are represented by color from red (for tape No. 1) to purple (for tape No. 40).

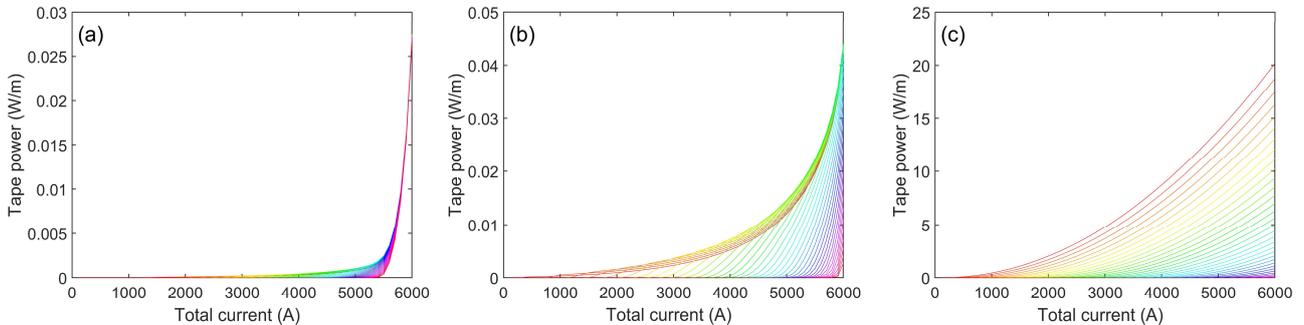

**Figure 6.** Joule heating power density in each tape at the outside cable part as a function of the total current in the case of (a) low-Rt, (b) mid-Rt, and (c) high-Rt. The 40 tapes are represented by color from red (for tape No. 1) to purple (for tape No. 40).

Figure 5 shows the voltage of different tapes in the outside cable part. In the low-Rt case, voltages in the 40 tapes have very modest differences, and they are almost coinciding when the total current is approaching the sum of tape $I_c$. Based on the 1 μV/cm criterion, these tape voltages would give the same cable $I_c$ about 5880 A, which is the sum of the individual tape $I_c$. For the mid-Rt case, the situation is again quite different. According to the same criterion, different tapes could give a quite different cable $I_c$ ranging from 5000 A to even more than 6000 A. For a high-Rt case, cable $I_c$ could be to the least 250 A if the voltage of tape No. 1 is monitored. On the contrary, if the voltage of tape No. 40 is used for measuring the cable $I_c$, one may find the cable shows no voltage but is burnt at a certain current point. As compared in figure 6, though the voltage difference in low and mid-Rt cases is already significantly different, the joule heating at 6000 A in the mid-Rt case is not that severe and is only about 1.5 times that in the low-Rt case. Instead, in the high-Rt case, the joule heating power increases by about three orders of magnitude at 6000 A.

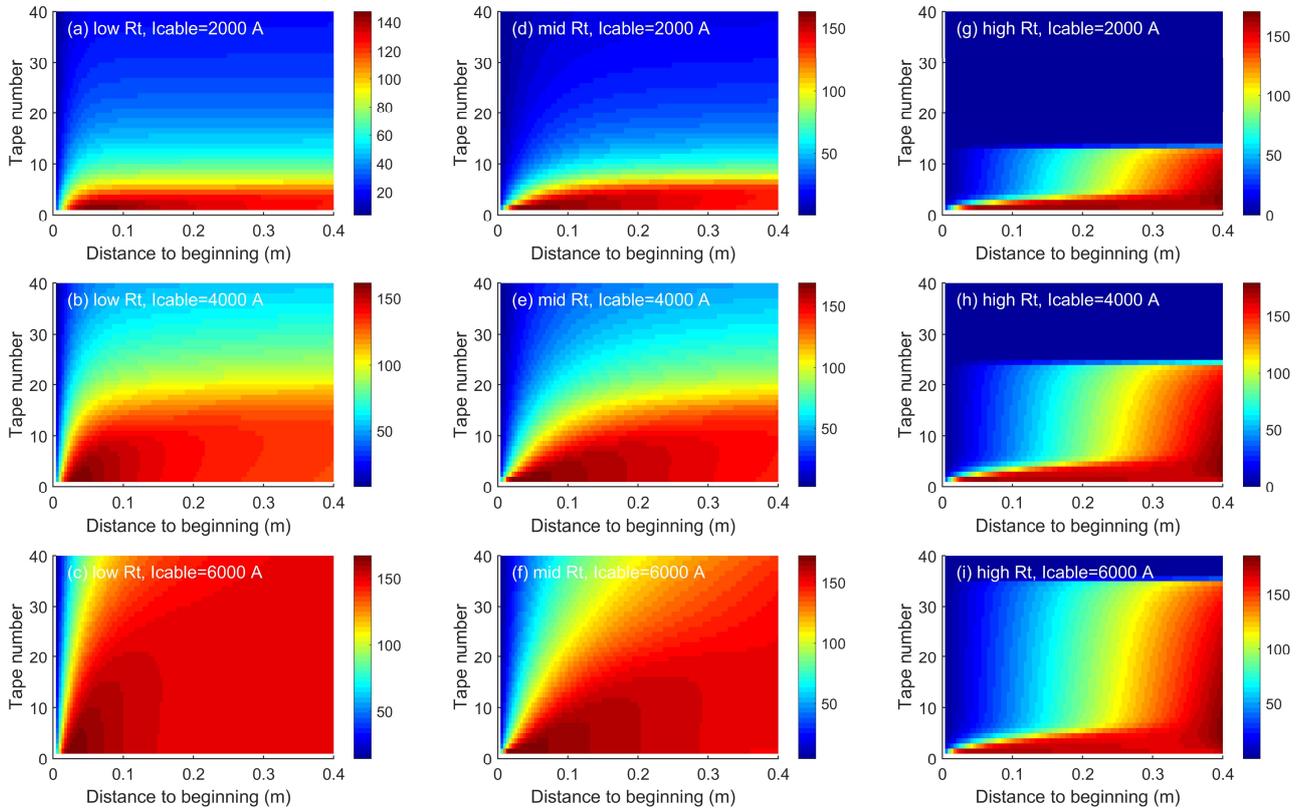

**Figure 7.** Current distribution among the tapes inside termination with different total currents and termination resistivities.

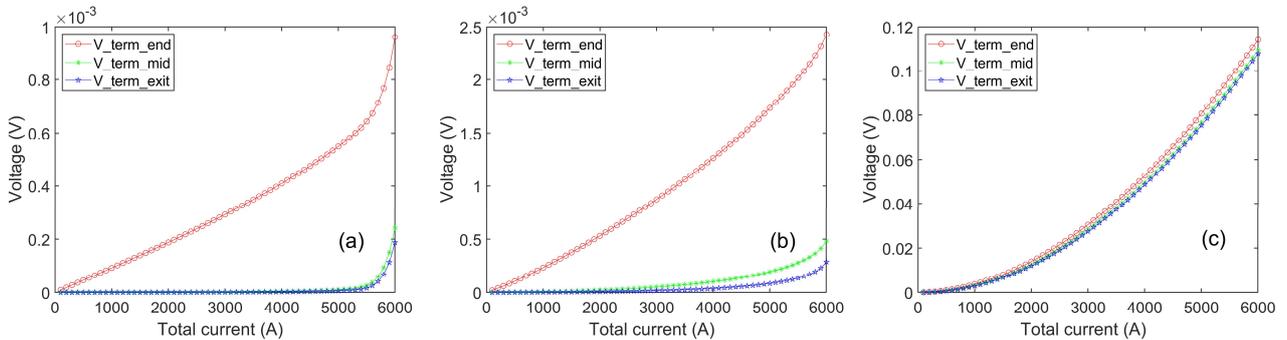

**Figure 8.** Voltage measured at various positions on the termination for (a) high-Rt, (b) mid-Rt, and (c) low-Rt.

As in these cases, current sharing only happens inside the copper termination, it is interesting to have a look at the current distribution among the tapes inside. Figure 7 shows the current distribution at one termination with different total currents for the three cases, which are significantly different due to the highly non-linear voltage-current (V-I) characteristic of REBCO tape. Despite the differences in pattern, one thing in common is that the local current in one tape could all exceed the tape $I_c$ around the current sharing frontiers, no matter the Rt. The difference is, the higher the Rt as well as its difference, the higher the maximum local current and it is closer to the exit. The occurrence of current exceeding $I_c$ locally is understandable: the current distribution is a result of minimizing the overall resistance or voltage drop. Imaging the current doesn't exceed tape $I_c$, then more current must bypass the contact resistance in a short length, in which route the voltage drop could be even higher.

A similar phenomenon is also observed with considering the current sharing among the cable in the outside part, which could further increase local tape current, as is discussed in section 3.3.

For the interest of sample testing, it is also important to know how the position of voltage pairs would influence the measured V-I characteristics of a cable, as figure 8 compares. Installing a voltage pair at the copper blocks is always the simplest. However, even in the case of low-Rt, if the voltage taps are installed at the middle or the end of the copper block (V_term_mid and V_term_end), a considerable voltage component contributed by the splice resistance will be measured and interferes with the V-I measurement. It should also be noted that this splice resistance contributed voltage has a non-linear relationship with the total current. In the case of mid-Rt and high-Rt, it is even impossible to distinguish a power-law transition from the voltages. As a result, the cable might be wrongly regarded as damaged with an incredibly low n-value. This behavior exists even when the voltage pair is close to the outside cable part (V_term_exit). In general, if the splice resistance is at a moderate level or worse, it would also be particularly important to carefully install the voltage taps. On the other hand, if the measured voltage of a tested cable increases at a small current with low n-values, it would be wise to check if the termination is well manufactured before questioning the cable performance.

3.3 The effect of current sharing among tapes

Six cases with different combinations of splice resistance and inter-tape resistance are simulated to study the influence of inter-tape current transfer on the current distribution of a whole cable, as listed in Table 2. First, the current distribution in tapes when the total current is 6000 A is compared for cases 4, 6, and 9, as shown in figure 9.

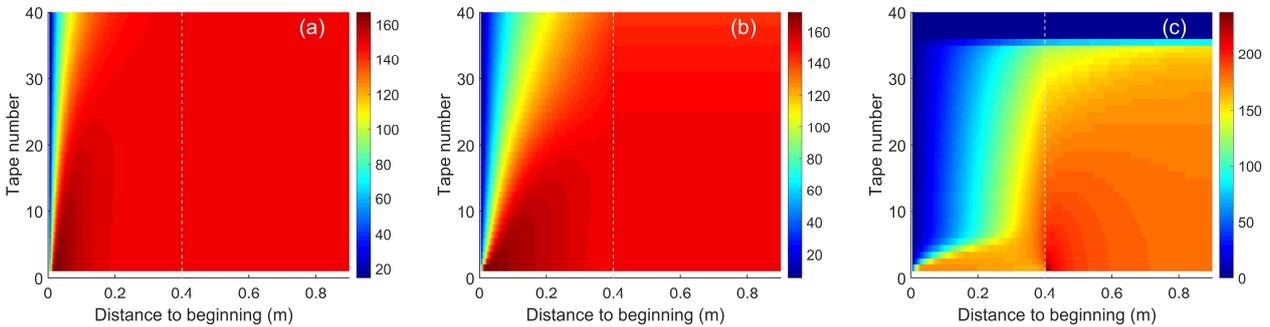

**Figure 9.** Current distribution among the tapes in the case of (a) low-Rt, (b) mid-Rt, and (c) high-Rt with a total current of 6000 A and $R_{cc}$ being 1000 nΩ·m. The figures contain one termination and half of the outside cable part. The white dash line indicates the termination exit.

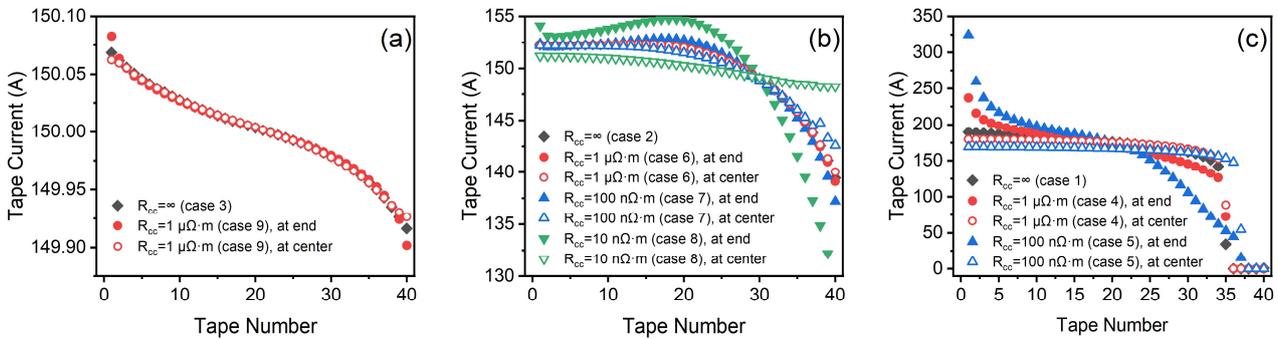

**Figure 10.** The influence of inter-tape current sharing on the current at the end or center of the outside cable part for (a) low-Rt, (b) mid-Rt, and (c) high-Rt.

For low and mid-Rt, as a rather balanced current distribution is already established through the termination and the $R_{cc}$ is relatively high, the inter-tape current sharing does not have notable change. For high-Rt, however, there is a counterintuitive result: the current distribution at the position that is just outside the termination (referred to as cable end) is further unbalanced. The maximum current in one tape greatly increases from 190 A to 237 A. Note the current at the middle of the cable (cable center) does slightly become more homogenized but the situation is not fundamentally improved. Figure 10 compares the influence of inter-tape resistance. For mid-Rt, it is until the $R_{cc}$ is low to a level of $1\times10^{-8}$ Ω·m, the inter-tape current sharing starts to have considerable influence on the current distribution, and local current increase at the cable end is also observed. For the high-Rt case, with better inter-tape current sharing, the highest current at the cable end is further increased to about 350 A. Of course, the overall distribution does become more average.

To briefly conclude the above results, when the tapes are not insulated from each other, the current sharing capacity will make the overall current distribution more balance, with a price of further increasing the maximum current at the cable end and could eventually burn out the cable when the splice resistance is in a poor condition. This is consistent with what we observed when testing short cable samples, the burning out always happens at the cable end.

3.4 Current sharing through additional copper layers

In most REBCO cable designs, a considerable amount of additional copper is added to reduce the quench current density. In this part, the effect of this additional copper on the current balance of a REBCO stack is also studied with several cases listed in Table 2. Referring to our first prototype of the X-cable, 4 layers of copper foil with a thickness of 0.1 mm are assumed to surround the stack but are simplified as one component in this calculation. Four cases with different splice resistances and contact resistances between the tape and additional copper are studied. Note that it is assumed all tapes have the same contact condition with the copper layer. In the first three cases, copper layers are assumed cut out near the termination and are not soldered into the copper block. There will be 3 cm (so 3 elements) at each end of the cable without copper layers covering it. In the last cases, the copper layers are regarded soldered into the copper block, but only with 1 cm.

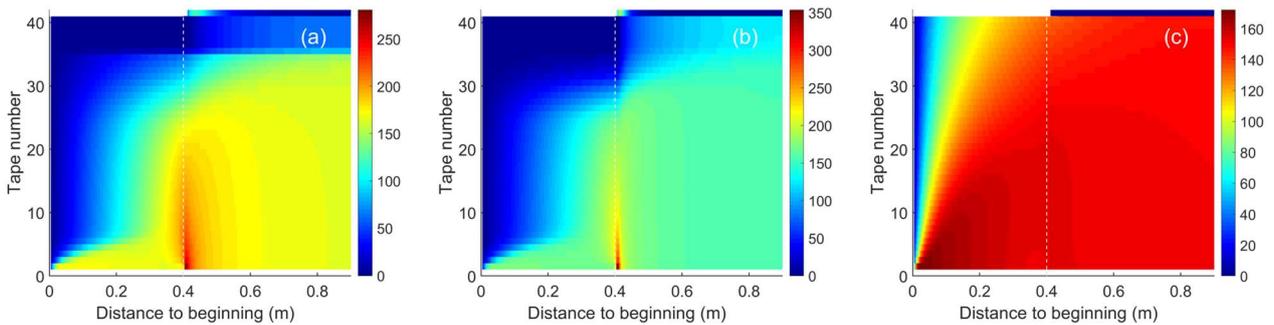

**Figure 11.** Current distribution among the tapes in cases 10-12 respectively. The figures contain one termination and half of the outside cable part. The white dash line indicates the termination exit. The copper layer is illustrated as tape No. 41 at the top.

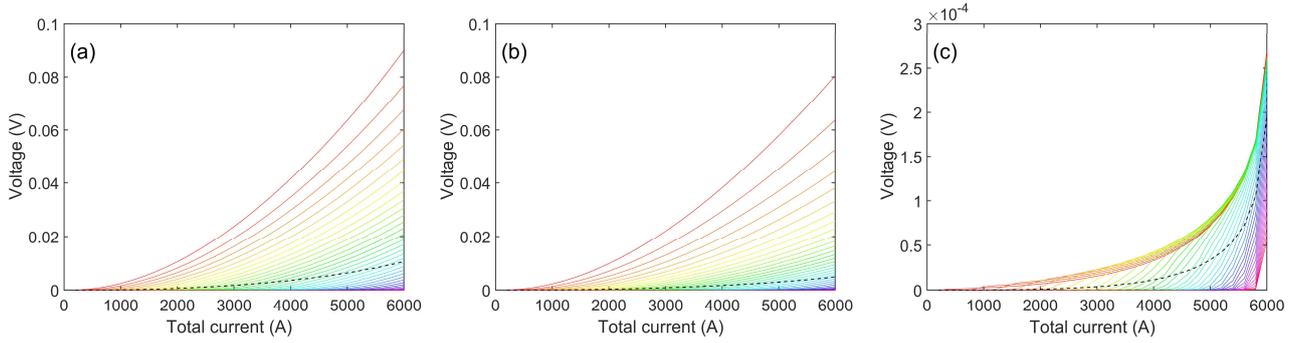

**Figure 12.** Voltage in each tape and copper layer at the outside cable part as a function of the total current for (a) case 10, (b) case 11, and (c) case 12. The 40 tapes are represented by color from red (for tape No. 1) to purple (for tape No. 40). The voltage in the copper layer is presented in a black dashed line.

Figure 11 compares the current distribution of the tapes and the additional copper. For the two high-Rt cases, compared with figure 9 the additional copper greatly helps to homogenize the current in different tapes even with a poor contact condition. However, the local current enrichment is also aggravated. The effect is even more significant than having a low inter-tape resistance. It should be noted that the copper layer itself does not take much longitudinal current and it mainly helps re-distributing the current, as it is assumed direct contact with each tape. Similar phenomena happen in the mid-Rt case.

In practice, it is also convenient to measure the voltage drop along the copper layer, so it would be interesting to see how different its behavior is compared with that in each tape. Figure 12 shows the V-I curves measured at the copper layer. Compared with the tape voltages shown in figure 5, the voltage in the copper layer is at a middle level. Note the tape voltages are only slightly reduced compared with cases 2 and 3.

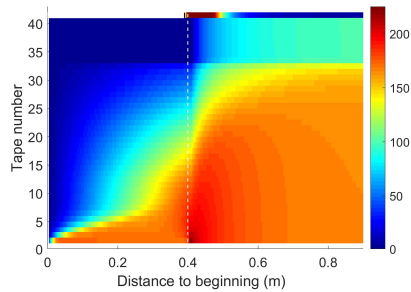

**Figure 13.** The current map at a total current of 6000 A in case 12.

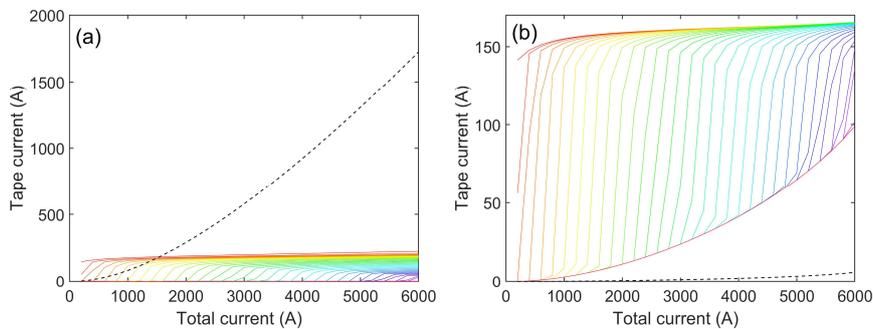

**Figure 14.** Current in different tapes at (a) the end, and (b) the center of the outside cable part.

If the copper layer is also soldered inside the termination even with a short length (~1 cm), things are already quite different. Figure 13 and 14 shows the current distribution. When the total current is 6000 A, more than a quarter of the total current pass through the copper layer and then re-distributes over the tapes. Note the cross-section area of the additional copper layer is more than 4 times that of the REBCO stack. The maximum current in one tape at the end of the cable decreases from 285 A to 225 A. This also verifies one of our observations that if the copper layer is soldered inside the termination, the cable is much more difficult to burn out.

## 4. The current unbalance in coils resulted from inductance difference.

4.1 The difference in inductance for stacked REBCO tapes

In practice, coil geometry defers a lot from one magnet to another. Here we simply have some calculations based on a hypothetical coil to have some general ideas. For a solenoid coil wound with a stack of 40 REBCO tapes, a reasonable radius would be larger than 0.1 m. The tape thickness is then much smaller than the coil radius. Consequently, the 40 tapes will have almost the same self-inductance (also see in [5,12]). The main difference would come from their mutual inductance matrix. Figure 15 shows the mutual inductances of two tapes at different distances and the total inductances (the sum of self-inductance and mutual inductances with all other tapes) of one tape at different positions. Three different coil radii of 0.1 m, 0.5 m, and 1 m are compared. Note the inductances are given in unit length calculated from a one-turn coil. For a multi-turn one, the absolute values will be higher according to the coil geometry. As expected, the tapes in the middle of a stack would endure higher total inductance. Consequently, when a time-varying current is charged to the cable, the current will first pass through the tapes at the top and bottom.

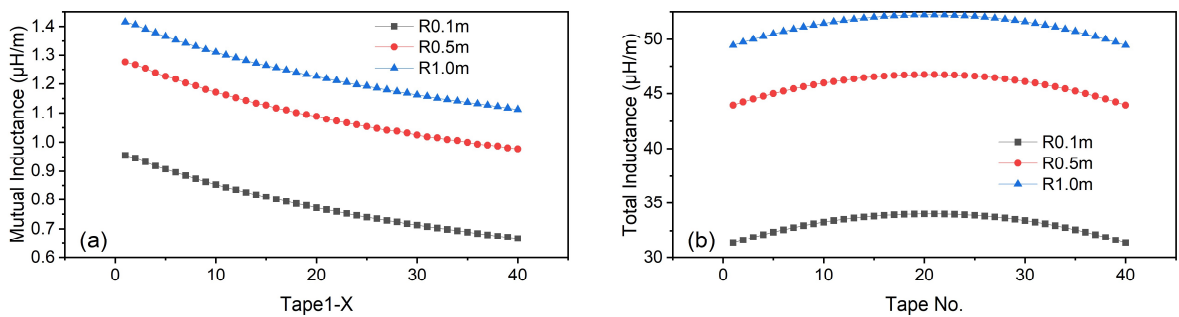

**Figure 15.** Mutual inductances of two tapes at (a) different distances, and (b) the total inductances of one tape at different positions, for different coil radii.

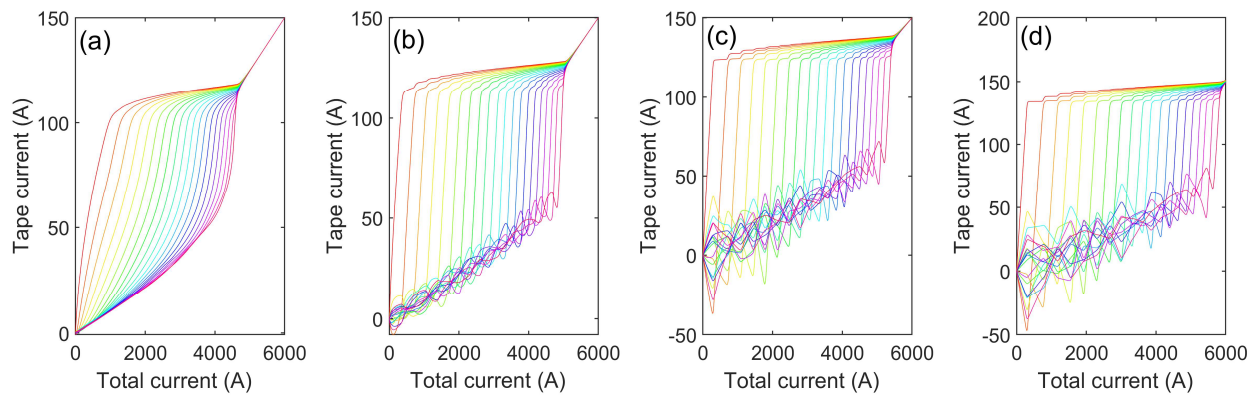

**Figure 16.** Current distribution in tapes at respectively (a) 1, (b) 10, (c) 100, and (d) 1000 A/s charging rate. The 20 tapes are represented by color from red (for tape No. 1) to purple (for tape No. 20)

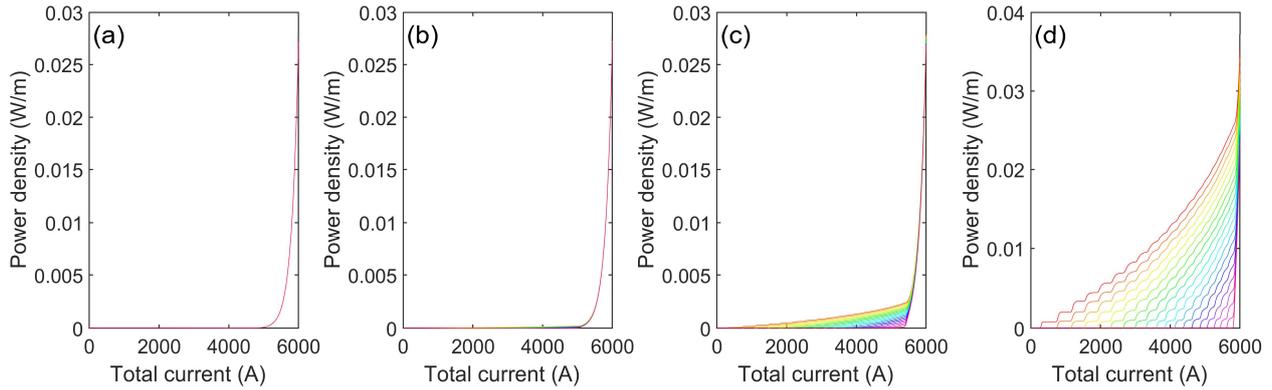

**Figure 17.** The Joule heating power density over the cable at respectively (a) 1, (b) 10, (c) 100, and (d) 1000 A/s charging rate. The 20 tapes are represented by color from red (for tape No. 1) to purple (for tape No. 20).

Figure 16 shows the current in each tape at different charging rates for a coil with a radius of 0.1 m and a total cable length of 100 m (so about 318 turns). Considering the symmetry, only the results of 20 tapes are presented. Current unbalance happens in all cases no matter the charging rate, but it is until the cable is charged with a very high rate of 1000 A/s then the unbalance keeps when the cable $I_c$ is reached. Otherwise, the resistive voltage will dominate the current distribution and homogenize it. It should be noted that the complex and changeable current distributions at the bottom of figure 16. (b)-(d) are not numerical instability, but are the effect of induction. At first, a certain current distribution is achieved to balance the inductive voltages. Then when the resistance in one tape gets countable, a new distribution must be established to reach a new balance. With so many tapes, each time of the re-balance of inductive voltages could greatly change the overall current distribution. Figure 17 shows the local joule heating of these cases. Note the AC loss is not considered here to simply compare the Joule heating effect resulting from unbalanced current distribution among tapes. Even in the fast-charging case, the maximum joule heating is only slightly increased. Do these results mean that the inductance mismatch induced current unbalance is not harmful to a REBCO stack?

In the above simulation, it is not only assumed $I_c$ of each tape is the same, but also the $I_c$ over the 100 m long tape is uniform. In practice, however, absolute uniformity over a long-length tape is usually not possible. In some magnet configurations, the local high field could also be responsible. If every REBCO tape has some short sections with low $I_c$, they might dominate the resistance in one tape. Figure 18 is the simulated results assuming the minimum $I_c$ in each cable is contributed by a 1 cm defect and the remaining part has uniformly distributed $I_c$ that is significantly higher (so never contributes to resistive voltage). As can be seen, in such a case, the current unbalance caused by inductance mismatch becomes much stronger. The joule heating, as shown in figure 19, also greatly increases. For a moderate charging rate of 10 A/s, the joule heating is already at a level of 1 W/m.

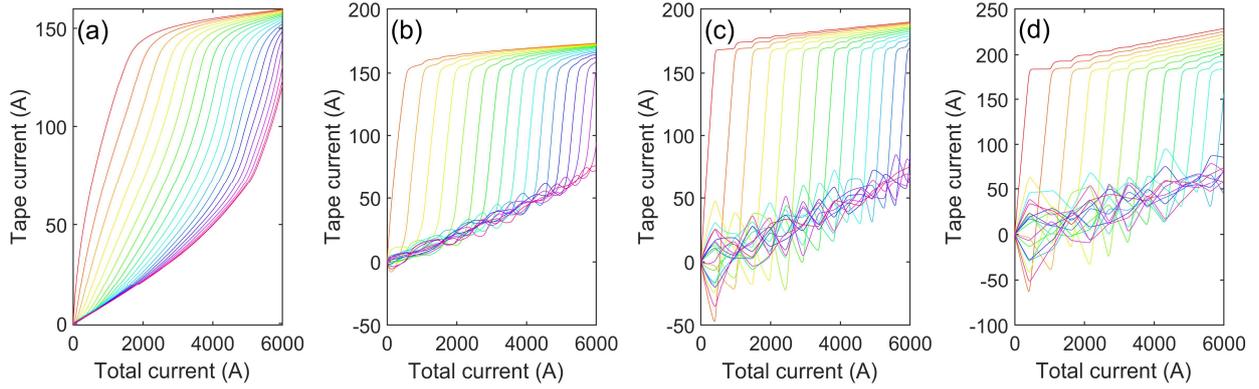

**Figure 18** Current distribution in tapes at respectively (a) 1, (b) 10, (c) 100, and (d) 1000 A/s charging rate assuming the tape resistance is contributed by a 1 cm local defect. The 20 tapes are represented by color from red (for tape No. 1) to purple (for tape No. 20)

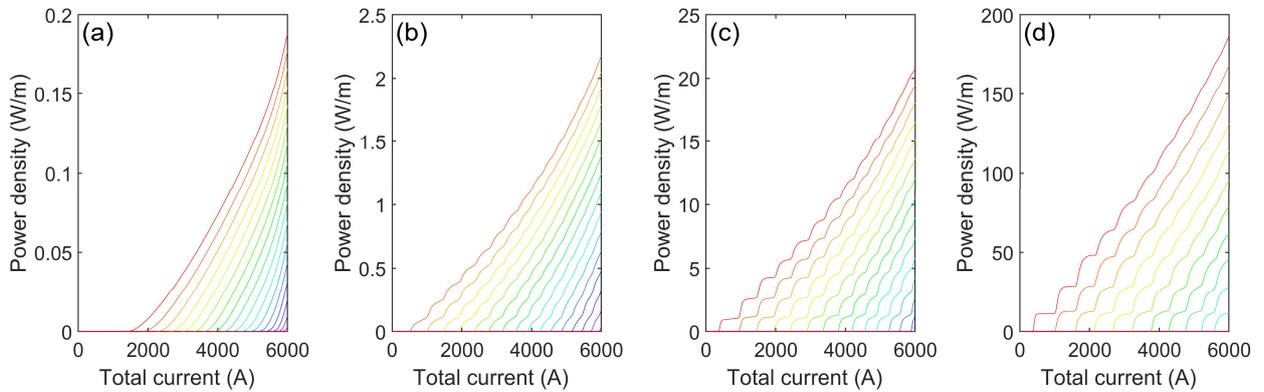

**Figure 19** The Joule heating power density at the local defects at respectively (a) 1, (b) 10, (c) 100, and (d) 1000 A/s charging rate assuming the tape resistance is contributed by a 1 cm local defect. The 20 tapes are represented by color from red (for tape No. 1) to purple (for tape No. 20).

A cure for the severe current unbalance caused by the combination of inductance mismatch and local defects is inter-tape current sharing. Unfortunately, a simulation for a 40-tape stack with 100 m long considering both inductance matrix and inter-tape current transfer is yet a mission-impossible for us because of the huge computing resources needed. Alternatively, we could have some estimations. Taking the example of the case with a 1000 A/s charging rate, the maximum current in one tape is about 224 A. To lower its current to 147 A, a current of 77 A must pass through the contact resistance with another tape and the voltage should be no more than 1 μV. If we wish the current transfer length to be 1 m, then the $R_{cc}$ must be less than $1.3 \times 10^{-8}$ Ω·m, which is a strict requirement that can only be fulfilled by soldering the stack together [16]. On the other hand, in a soldered stack, one tape could share its current with almost any other tape, so this kind of current unbalance is unlikely a problem. For a slow charging rate of 1 A/s, the required $R_{cc}$ decreases to $1 \times 10^{-7}$ Ω·m, which is a level that can barely be accessed by good direct contact.

## 5. Summary

The current unbalance in stacked REBCO tapes that are resulted from differences in splice resistances and inductance matrix, both of which are related to the non-transposed nature of stacked REBCO tapes, are numerically studied with an electrical circuit grid model.

For short sample testing, the splice resistivity for each tape should be at a magnitude of 10 nΩ·m with the difference among tapes less than 1 nΩ·m to totally avoid splice resistance interfering with the measurement of a REBCO cable. A moderate range from 10 to 1000 nΩ·m would already significantly change a cable's *V-I* characteristic and make a cable's $I_c$ underestimated. If the measured voltage of a REBCO cable increases at a very small current with low *n*-values, it would be wise to check if the termination is not well manufactured before questioning the cable performance. For this kind of current unbalance, the inter-tape current sharing is found not truly helping. The current distribution at the cable end is more unbalanced, which is extremely dangerous if the termination is not well manufactured. The copper protection layer could also worsen the situation since it can also help share currents among tape. However, if it is also soldered inside the termination (or may be connected by other methods) then it could greatly release the danger.

The different inductances of the stacked REBCO tapes also cause current unbalance during charging. If the $I_c$ is perfectly uniform over the tape length, then this unbalance is unlikely a problem regarding the cable's electrical performance even for a high charging rate of 1 kA/s. However, when considering the practical $I_c$ variation over tape length, good inter-tape current sharing capability should be ensured to minimize the effect. For example, the inter-tape resistivity should be as low as $1.3 \times 10^{-8}$ Ω·m when local defects exist to make sure a charging rate of 1 kA/s is not harmful concerning Joule heating.

It is worth mentioning that this manuscript focuses on the unbalanced current and the resulting electrical characteristics around the cable's $I_c$. Another important consequence that is not discussed above is the field quality (uniformity). According to the simulated results, strong current unbalance is almost inevitable for stacked REBCO tapes at currents far from its total critical current, which is especially complex during fast charging. As a result, cable concepts based on stacked REBCO tapes should be carefully used in applications emphasizing field quality, or the influence, possibly together with the screening current in a tape, should be analyzed in advance.

# 6. Reference

[1] Bruzzone P 2006 30 Years of Conductors for Fusion: A Summary and Perspectives *IEEE Transactions on Applied Superconductivity* **16** 839–44

[2] Bruzzone P 2015 Superconductivity and fusion energy—the inseparable companions *Supercond. Sci. Technol.* **28** 024001

[3] Uglietti D 2019 A review of commercial high temperature superconducting materials for large magnets: from wires and tapes to cables and conductors *Supercond. Sci. Technol.* **32** 053001

[4] Goldacker W, Frank A, Kudymow A, Heller R, Kling A, Terzieva S and Schmidt C 2009 Status of high transport current ROEBEL assembled coated conductor cables *Supercond. Sci. Technol.* **11**

[5] Takayasu M, Chiesa L, Bromberg L and Minervini J V 2012 HTS twisted stacked-tape cable conductor *Supercond. Sci. Technol.* **25** 014011

[6] Uglietti D, Bykovsky N, Wesche R and Bruzzone P 2015 Development of HTS Conductors for Fusion Magnets *IEEE Transactions on Applied Superconductivity* **25** 1–6

[7] Augieri A, De Marzi G, Celentano G, Muzzi L, Tomassetti G, Rizzo F, Anemona A, Bragagni A, Seri M, Bayer C, Bagrets N and della Corte A 2015 Electrical Characterization of ENEA High Temperature

Superconducting Cable *IEEE Transactions on Applied Superconductivity* **25** 1–4

[8] Wolf M J, Fietz W H, Bayer C M, Schlachter S I, Heller R and Weiss K-P 2016 HTS CroCo: A Stacked HTS Conductor Optimized for High Currents and Long-Length Production *IEEE Transactions on Applied Superconductivity* **26** 19–24

[9] Yanagi N, Terazaki Y, Narushima Y, Onodera Y, Hirano N, Hamaguchi S, Chikaraishi H, Takada S, Ito S and Takahata K 2022 Progress of HTS STARS Conductor Development for the Next-Generation Helical Fusion Experimental Device *Plasma and Fusion Research* **17** 2405076–2405076

[10] Hartwig Z S, Vieira R F, Sorbom B N, Badcock R A, Bajko M, Beck W K, Castaldo B, Craighill C L, Davies M, Estrada J, Fry V, Golfinopoulos T, Hubbard A E, Irby J H, Kuznetsov S, Lammi C J, Michael P C, Mouratidis T, Murray R A, Pfeiffer A T, Pierson S Z, Radovinsky A, Rowell M D, Salazar E E, Segal M, Stahle P W, Takayasu M, Toland T L and Zhou L 2020 VIPER: an industrially scalable high-current high-temperature superconductor cable *Supercond. Sci. Technol.* 9

[11] Wang J, Kang R, Chen X, Yang C, Wang Y, Wang C and Xu Q 2022 Development of a Roebel structure transposed cable with in-plane bending of REBCO tapes *Superconductivity* **3** 100019

[12] Uglietti D, Kang R, Wesche R and Grilli F 2020 Non-twisted stacks of coated conductors for magnets: Analysis of inductance and AC losses *Cryogenics* **110** 103118

[13] Cau F and Bruzzone P 2009 Inter-strand resistance measurements in the termination of the ITER SULTAN samples *Supercond. Sci. Technol.* **22** 045012

[14] Takayasu M, Chiesa L and Minervini J V 2016 *Termination Methods for REBCO Tape High-Current Cable Conductors*

[15] De Marzi G, Celentano G, Augieri A, Marchetti M and Vannozzi A 2021 Experimental and numerical studies on current distribution in stacks of HTS tapes for cable-in-conduit-conductors *Supercond. Sci. Technol.* **34** 035016

[16] Bykovsky N 2017 *HTS high current cable for fusion application* (EPFL)

[17] Kim J-H, Kim C H, Pothavajhala V and Pamidi S V 2013 Current Sharing and Redistribution in Superconducting DC Cable *IEEE Transactions on Applied Superconductivity* **23** 4801304–4801304

[18] Zermeno V, Krüger P, Takayasu M and Grilli F 2014 Modeling and simulation of termination resistances in superconducting cables *Supercond. Sci. Technol.* **27** 124013

[19] Willering G P, van der Laan D C, Weijers H W, Noyes P D, Miller G E and Viouchkov Y 2015 Effect of variations in terminal contact resistances on the current distribution in high-temperature superconducting cables *Supercond. Sci. Technol.* **28** 035001

[20] Takayasu M, Chiesa L, Allen N C and Minervini J V 2016 Present Status and Recent Developments of the Twisted Stacked-Tape Cable Conductor *IEEE Transactions on Applied Superconductivity* **26** 25–34

[21] Bottura L, Rosso C and Breschi M 2000 A general model for thermal, hydraulic and electric analysis of

superconducting cables *Cryogenics* **40** 617–26

[22]     Bonura M, Barth C, Joudrier A, Troitino J F, Fete A and Senatore C 2019 Systematic Study of the Contact Resistance Between REBCO Tapes: Pressure Dependence in the Case of No-Insulation, Metal Co-Winding and Metal-Insulation *Ieee Transactions on Applied Superconductivity* **29** 6600305

[23]     Dicuonzo O 2022 *Electromechanical investigations and quench experiments on sub-size HTS cables for high field EUDEMO Central Solenoid* (EPFL)